\documentstyle[12pt]{article}
\def\PRL #1 #2 #3 {Phys.~Rev.~Lett.~{\bf #1}, #2 (#3)}
\def\PRD #1 #2 #3 {Phys.~Rev.~D~{\bf #1}, #2 (#3)}
\def\PLB #1 #2 #3 {Phys.~Lett.~{\bf B#1}, #2 (#3)}
\def\NPB #1 #2 #3 {Nucl.~Phys.~{\bf B#1}, #2 (#3)}

\newcommand{\gtap}{{\raise.3ex\hbox{$>$\kern-.75em\lower1ex\hbox{$\sim$}}}}
\newcommand{\ltap}{{\raise.3ex\hbox{$<$\kern-.75em\lower1ex\hbox{$\sim$}}}}

\begin{document}
\include{psfig}
\begin{titlepage}

\rightline{hep-ph/9804227}
\rightline{UCSD/PTH 98-13}
\rightline{ILL-(TH)-98-2}
\medskip
\bigskip\bigskip
\begin{center}
{\Large \bf Quarkonia and the Pole Mass} \\
\medskip
\bigskip\bigskip\bigskip
{\large \bf A.~H.~Hoang} \\
\medskip
Department of Physics \\
University of California, San Diego \\ 9500 Gilman Drive \\
La Jolla, CA\ \ 92093-0319 \\
\bigskip\bigskip
{\large{\bf M.~C.~Smith, T.~Stelzer} and 
{\bf S.~Willenbrock}} \\ 
\medskip 
Department of Physics \\
University of Illinois \\ 1110 West Green Street \\  Urbana, IL\ \ 61801 \\
\bigskip 
\end{center} 
\bigskip\bigskip\bigskip

\begin{abstract}
The pole mass of a heavy quark is ambiguous by an amount of order 
$\Lambda_{QCD}$.  We show that the heavy-quark potential, $V(r)$,
is similarly ambiguous, but that the total static energy, $2M_{pole}+V(r)$, is
unambiguous when expressed in terms of a short-distance mass.  This implies
that the extraction of a short-distance mass from the quarkonium spectrum
is free of an ambiguity of order $\Lambda_{QCD}$, in contrast with the pole 
mass.
\end{abstract}

\end{titlepage}

\newpage

The pole mass of a heavy quark is known to be well defined at any finite 
order in 
perturbation theory \cite{K}.  However, it is known that the pole mass
of a heavy quark is an ambiguous concept at large orders 
in perturbation theory \cite{BU,BSUV,BB}.  The perturbative series relating
the pole mass to a short-distance mass (such as the $\overline{\rm MS}$ mass)
has coefficients which grow factorially, leading to an ambiguity of order 
$\Lambda_{QCD}$ in the pole mass \cite{BSUV,BB}.  
This factorial divergence is related to the existence
of an infrared renormalon pole \cite{tH} in the Borel-transformed quark 
self-energy when evaluated on shell \cite{BB}.

The ambiguity in the heavy-quark pole mass can be understood heuristically
in terms of confinement.  The pole mass of a heavy meson 
(bound state of a heavy quark and a light\footnote{$m<\Lambda_{QCD}$} 
antiquark) is well defined, since
it is a physical quantity.  To extract the pole mass of the heavy quark
from the meson mass, one must subtract the binding energy, 
of order $\Lambda_{QCD}$.  However, it is impossible to define
this binding energy unambiguously, since one cannot separate the 
light antiquark from
the heavy quark, due to confinement.  Hence the pole mass of a heavy quark
is ambiguous by an amount of order $\Lambda_{QCD}$.

Consider instead quarkonium, a bound state of a heavy quark and a heavy
antiquark.  If the mass of the heavy quark is sufficiently large, 
the strong coupling governing the heavy-quark interactions in
quarkonia states with small principal quantum number is weak \cite{AP}.
The quarkonium can then be handled just like
positronium in electrodynamics, by summing uncrossed Coulomb ladder
diagrams to yield the Schr\"odinger equation, and using non-relativistic
perturbation theory.
Thus the low-lying quarkonia states can be described almost 
entirely in terms of QCD perturbation theory.
It is tempting to conclude that,
if the quark mass is heavy enough,
the pole mass can be extracted to arbitrary precision from a 
perturbative calculation of the binding 
energy of these states, perhaps 
supplemented by some nonperturbative input. 
The goal of this article is to show
that this is not the case; the heavy-quark pole mass cannot be 
extracted from the quarkonium spectrum with an accuracy better than order
$\Lambda_{QCD}$.  This is an important point to clarify, since the 
$c$- and $b$-quark pole masses 
are extracted from the $J/\psi$ \cite{G,N,TY,DGP,FNAL} and $\Upsilon$ spectra 
\cite{G,N,TY,NRQCD,V,JP,KPP,H}.

We begin by considering the static potential between a heavy quark and a
heavy antiquark in a color singlet state
at leading order in QCD \cite{S,F,ADM,Fein,P},
\begin{equation}
V(r) = -C_F4\pi\alpha_s\int\frac{d^3{\bf k}}{(2\pi)^3}
\frac{e^{i{\bf k}\cdot{\bf r}}}{{\bf k}^2} 
= -C_F\frac{\alpha_s}{r} \;\;\;(C_F = 4/3)\;
\,,
\label{pot}
\end{equation}
where $\alpha_s \equiv \alpha_s(\mu)$ is the strong coupling evaluated at the 
scale $\mu$.
An elegant means to analyze large orders in
perturbation theory is to calculate the Borel transform (with respect to 
$b_0\alpha_s/4\pi$) of the static
potential.  This calculation, performed in Ref.~\cite{AL}, 
gives\footnote{An analysis of quarkonia energies at large orders in 
perturbation theory based on the position-space potential 
$V(r) = -C_F\frac{\alpha_s(1/r)}{r}$ may be found in Ref.~\cite{SSW}.} 
\begin{eqnarray}
\tilde V(r) & = & - C_F\frac{(4\pi)^2}{b_0}\left(\frac{e^C}{\mu^2}\right)^{-u}
\int\frac{d^3{\bf k}}{(2\pi)^3}
\frac{e^{i{\bf k}\cdot{\bf r}}}{{\bf k}^{2(1+u)}} \nonumber \\
& = & -C_F\frac{4}{b_0}e^{-Cu}\frac{1}{r}(\mu r)^{2u}
\frac{\Gamma(\frac{1}{2}+u)\Gamma(\frac{1}{2}-u)}{\Gamma(2u+1)} \nonumber \\
& = & C_F\frac{4}{b_0}e^{-C/2}\mu\frac{1}{(u-\frac{1}{2})} + \cdots 
\,,
\label{vborel}
\end{eqnarray}
where $u$ is the Borel parameter, $C$ is a renormalization-scheme-dependent
constant ($C=-5/3$ in the $\overline{\rm MS}$ scheme), and
\begin{equation}
b_0\equiv 11-\frac{2}{3}N_f
\end{equation}
is the one-loop beta-function coefficient.  The infrared renormalon pole
nearest the origin, at $u=1/2$, controls the asymptotic behavior of the 
perturbation series.  This pole and its residue are made explicit in the last
line of Eq.~(\ref{vborel}).  

The static potential is recovered from Eq.~(\ref{vborel})
by inverse Borel transformation,
\begin{equation}
V(r) = \int_0^\infty du\,e^{-4\pi u/(b_0\alpha_s)}\tilde V(r)\;.
\end{equation}
The evaluation of this integral is impeded by the presence of infrared 
renormalon poles in $\tilde V(r)$ on the positive real axis at all 
half-integer values of $u$.
The ambiguity in the integral is dominated by the infrared renormalon pole
closest to the origin, at $u=1/2$.  Estimating the ambiguity as half the 
magnitude of the
difference between deforming the integration contour above and below the pole
yields
\begin{eqnarray}
\delta V(r) & \sim & C_F\frac{4\pi}{b_0}e^{-C/2}\mu
e^{-2\pi/(b_0\alpha_s)} 
\nonumber \\
& \sim & C_F\frac{4\pi}{b_0}e^{-C/2}\Lambda_{QCD}
\,,
\end{eqnarray}
where the one-loop renormalization-group equation for $\alpha_s$ has been
used to
obtain the final expression.  The ambiguity is a constant shift of the 
potential, by an amount of order $\Lambda_{QCD}$. 

The ambiguity of order $\Lambda_{QCD}$ in the static potential was first
derived in Ref.~\cite{AL}, but no attempt was made to interpret it.  We now 
show that this ambiguity is due to the (implicit) use of the pole mass,
which is sensitive to momenta of order $\Lambda_{QCD}$,
in the definition of the static potential.  

The relation between the heavy-quark 
pole mass, $M_{pole}$, and a short-distance mass, $M$, is given in 
Borel space by \cite{BB}
\begin{eqnarray}
\tilde M_{pole} & = & M\left(\delta(u)
+\frac{C_F}{b_0}\left[\left(\frac{M^2}{\mu^2}\right)^{-u}e^{-Cu}6(1-u)
\frac{\Gamma(u)\Gamma(1-2u)}{\Gamma(3-u)}
-\frac{3}{u}+R_{\Sigma_1}(u)\right]\right) \nonumber \\
& = & -C_F\frac{2}{b_0}e^{-C/2}\mu\frac{1}{(u-\frac{1}{2})} + \cdots
\,,
\label{mpole}
\end{eqnarray}
where $R_{\Sigma_1}(u)$ is a renormalization-scheme-dependent regular function.
The infrared renormalon pole closest to the origin is again at $u=1/2$,
and its residue is displayed in the last line of Eq.~(\ref{mpole}).
The ambiguity in the pole mass 
can be estimated as above for the static potential,
and is given by \cite{BB}
\begin{equation}
\delta M_{pole} \sim C_F\frac{2\pi}{b_0}e^{-C/2}\Lambda_{QCD} \,,
\end{equation}
as is well known.

Consider now the total static energy of a heavy quark and a heavy
antiquark in a color singlet state
with spatial separation $r$,
given by the sum of the static potential energy and the rest 
mass of the particles.  In Borel space, 
\begin{equation}
\tilde E_{static}(r) = 2\tilde M_{pole} + \tilde V(r)\;.
\label{estatic}
\end{equation}
If the pole mass is eliminated in favor of a short-distance mass via 
Eq.~(\ref{mpole}), we see that the infrared renormalon pole at $u=1/2$
in the total static energy is cancelled.  Thus the total static energy does
not have an ambiguity of order $\Lambda_{QCD}$ when expressed in terms of
a short-distance mass.

The total static energy of the heavy quark-antiquark pair in a 
color-singlet state is an unambiguous concept 
\cite{Wilson,S}. However, when one attempts to separate this energy 
into the sum of the static potential energy and the pole masses of the quark 
and antiquark \cite{S,F,ADM,Fein,P}, the ambiguity of order
$\Lambda_{QCD}$ in the pole mass 
results in  
a corresponding ambiguity in the static potential.  This is the
source of the ambiguity of order $\Lambda_{QCD}$ in the static
potential. The 
ambiguity disappears when the pole mass is replaced with a short-distance 
mass.  This well-known phenomenon occurs for many other processes in QCD 
\cite{BSUV,BB,BBZ,NNA,SV,GS,LMS,NS}.

Because it will be useful to us when we consider the full dynamical 
(nonstatic) quarkonium calculation, let us recall that an analysis of the
infrared behavior can also be carried out via a finite gluon mass
$\lambda$ \cite{BSUV,BBZ,Ball1}.  The static potential becomes
\begin{equation}
V(r) = - C_F\frac{\alpha_s}{r}e^{-\lambda r} 
= -C_F\alpha_s\left(\frac{1}{r}-\lambda + \cdots\right)\;.
\label{vlam}
\end{equation}
The term of order $\lambda$ corresponds to the 
infrared renormalon pole
at $u=1/2$. The relation between the pole 
mass and a short-distance mass is \cite{BSUV,BBZ,Ball1}
\begin{equation}
M_{pole} = M - C_F\frac{\alpha_s}{2}\lambda + \cdots\;.
\label{mlam}
\end{equation}
We see that the term linear in the gluon mass cancels when we calculate 
the total static energy, $2M_{pole}+V(r)$, in terms of a short-distance mass.
This corresponds to the cancellation of the pole at $u=1/2$ in the Borel 
transform of the total static energy, Eq.~(\ref{estatic}). 

The full dynamical (nonstatic) quarkonium calculation
at leading order in the nonrelativistic approximation 
requires solving the Schr\"odinger equation
\begin{equation}
\left(-\frac{\nabla^2}{M_{pole}} + V(r) - E \,\right)
G({\bf r},0,E) 
= \delta^{(3)}({\bf r})\;,
\label{schr}
\end{equation}
where $G({\bf r},0,E)$ is the Schr\"odinger-equation Green function,
and $E\equiv \sqrt s - 2 M_{pole}$ is the 
binding energy.  The total center-of-mass energy $\sqrt s$
is physical and unambiguous, but the binding energy $E$ is not, 
since it requires subtracting twice
the pole mass.  
Eliminating $E$ in favor of $\sqrt s$ in Eq.~(\ref{schr}) gives
\begin{equation}
\left(-\frac{\nabla^2}{M_{pole}} + 2M_{pole} + V(r) - \sqrt s \,\right)
G({\bf r},0,\sqrt s-2M_{pole}) 
= \delta^{(3)}({\bf r})\;.
\label{schr2}
\end{equation}
The quarkonia masses correspond to the poles of the Green function 
in the $\sqrt s$ plane. 
We see that the total static energy, $2M_{pole}+V(r)$, now appears in the 
Schr\"odinger equation.  As shown 
above, the total static energy is not ambiguous 
by an amount of order $\Lambda_{QCD}$,
but due to the ambiguity in $V(r)$, the pole mass cannot be 
extracted from the quarkonium spectrum with an accuracy better than order
$\Lambda_{QCD}$.  However, the static potential can be made free of the
ambiguity of order $\Lambda_{QCD}$ if the pole mass is exchanged for a 
short-distance mass and if twice the difference between the pole and the
short-distance mass is absorbed into the static potential,
\begin{equation}
\left(-\frac{\nabla^2}{M_{pole}} + 2M + \hat V(r) - \sqrt s \,\right)
\hat G({\bf r},0,\sqrt s-2M) 
= \delta^{(3)}({\bf r})\;,
\label{schr3}
\end{equation}
where
\begin{equation}
\hat V(r) = V(r)+2(M_{pole}-M)\;.
\end{equation}
Hence the accuracy with which
a short-distance mass can be extracted from the quarkonium spectrum is not
limited by order $\Lambda_{QCD}$.  

The pole mass also appears 
in the denominator of the kinetic-energy term in the Schr\"o\-din\-ger 
equation. 
However, replacing this mass with a short-distance mass 
only affects terms suppressed by powers of $\alpha_s$.  This
can be seen most easily
by using the gluon mass as an infrared regulator, 
Eqs.~(\ref{vlam}) and (\ref{mlam}).  The linear 
gluon mass terms which cancel in the total static energy are of order 
$\alpha_s\lambda$.
Since the kinetic-energy term is of order $M_{pole}\alpha_s^2$, the linear 
gluon mass term generated by replacing $M_{pole}$ with a short-distance mass 
is of order $\alpha_s^3\lambda$. 

In this paper we have shown that the heavy-quark pole mass cannot be 
extracted from 
the quarkonium spectrum with an accuracy better than order $\Lambda_{QCD}$.
This is relevant for the determination of the $c$- and $b$-quark pole masses 
from the $J/\psi$ \cite{G,N,TY,DGP,FNAL} and $\Upsilon$ spectra 
\cite{G,N,TY,NRQCD,V,JP,KPP,H}. 
However, the accuracy with which a properly-defined short-distance
mass can be extracted from the quarkonium spectrum is not limited by order
$\Lambda_{QCD}$.

{\it Note added}: The results of this paper have also been 
arrived at in a recent paper \cite{B}.

\section*{Acknowledgments}

\indent\indent We are grateful for conversations with A.~El-Khadra. 
A.~H.~H.~was supported in part by the U.~S.~Department of Energy under
contract No.~DOE~DE-FG03-90ER40546.
M.~S., T.~S., and S.~W.~were supported in part by the U.~S.~Department 
of Energy under contract No.~DOE~DE-FG02-91ER40677.  M.~S.~was supported 
in part by a grant from the UIUC Campus Research Board.

\end{document}